\documentclass[conference]{IEEEtran}
\pdfoutput=1
\usepackage{graphicx}
\usepackage{balance}
\usepackage{amsthm}
\usepackage[english]{babel}

\begin{document}
\IEEEspecialpapernotice{A version of this work is in press for the proceedings of 7th IEEE International Workshop on Information Forensics and Security (WIFS) 2015. 
IEEE Copyright Notice:  Personal use of this material is permitted. Permission from IEEE must be obtained for all other uses, in any current or future media, including reprinting/republishing this material for advertising or promotional purposes, creating new collective works, for resale or redistribution to servers or lists, or reuse of any copyrighted component of this work in other works.}
\title{TRAP: using TaRgeted Ads to unveil Google personal Profiles}
\author{\IEEEauthorblockN{Mauro Conti}
\IEEEauthorblockA{University of Padua,\\
Padua, Italy\\
Email: conti@math.unipd.it}
\and
\IEEEauthorblockN{Vittoria Cozza, Marinella Petrocchi, Angelo Spognardi}
\IEEEauthorblockA{Institute of Informatics and Telematics of CNR\\
 Pisa, Italy\\
Email: {\{name.surname\}@iit.cnr.it}}
}

\maketitle              

\begin{abstract}
  In the last decade, the advertisement market spread significantly in
  the web and mobile app system. Its effectiveness is also due thanks
  to the possibility to target the advertisement on the specific
  interests of the actual user, other than on the content of the
  website hosting the advertisement. In this scenario, became of great
  value services that collect and hence can provide information about
  the browsing user, like Facebook and Google.
  In this paper, we show how to maliciously exploit the Google
  Targeted Advertising system to infer personal information in Google
  user profiles. In particular, the attack we consider is external
  from Google and relies on combining data from Google AdWords with
  other data collected from a website of the Google Display Network.
  We validate the effectiveness of our proposed attack, also
  discussing possible application scenarios. The result of our
  research shows a significant practical privacy issue behind such
  type of targeted advertising service, and call for further
  investigation and the design of more privacy-aware solutions,
  possibly without impeding the current business model involved in
  online advertisement.
\end{abstract}
\section{Introduction}\label{sec:introduction}
Online advertisement (also referred as ``ad" in the following) generates a business of
hundreds of billions dollars.\footnote{Techcrunch.com: Internet Ad Spend To Reach \$121B In 2014, 23\% Of \$537B Total Ad Spend, Ad Tech Boosts Display: {\it on.tcrn.ch/l/qsMM}}
Online ad companies help advertisers to reach the
best possible consumers, i.e.,  those users that would be more
sensible to a specific type of ads. Google is currently
dominating the online ad market, thanks to 
two dedicated  services: AdSense\footnote{{\it www.google.com/adwords/}}, through which everyone can easily create advertisements. 

The Google ads framework involves three main actors: the {\it advertiser}, the {\it  publisher},
and the {\it customer}. The customer is every user navigating the Web and seeing online ads. The advertiser wants to reach the customer, in order to show her ads, the publisher sells a space on her website to the advertiser: ads are shown to the visitors of that website.
In the whole process, Google acts as a mediator between advertisers and publishers,
telling which advertisement has to be shown on which website,
depending on the type of visitors that website has. 
Furthermore, the advertiser can specify
the type of audience to reach, achieving the so called
\textit{targeted advertising}. Targeted advertising has the double
advantage (when compared to  non-targeted ones) to be interesting for the advertiser and---at the same time---pleasant
for the customer, which receives ads closer to her interest. Nonetheless, the Google targeted advertising system is exploitable to reveal
personal information of the customers: knowing that a
specific ad has been shown to a user may simply reveal her
interests~\cite{castelluccia2012,icdmw2010}. 

\noindent\textit{Contributions:} In this paper, we design and test a novel attack to
discover personal data (such as users interests, stored in the Google
ads users' profile), based on ad
impressions.
Our approach is particularly novel in the fact that we can reconstruct
the user profile ``remotely", i.e., without having any direct access
to the advertisements the user sees.

\noindent\textit{Roadmap:} In the following section, we present the Google targeted advertising system. Section~\ref{sec:TRAP:-our-system} describes the attack that can be remotely launched to discover the users' profile. In Section~\ref{sec:victim-profiles},  we show our
experimental results that confirm the feasibility of the proposal. Section~\ref{sec:related-works} discusses related work in the area. Section~\ref{sec:conclusions} gives  final remarks.

\section{The Google targeted advertising system}\label{sec:google-targ-advert}

Google offers a complex and rich system for targeted advertising, through the so called DoubleClick infrastructure. This can manage several types
of ads: display-based ads (located within pages of
websites), search-based ads (which appear among the results of a Google search),
ads in the YouTube platform and ads in the Gmail service. In this work, we will focus on display-based ads. 

The two main components of the Google targeted advertising system are AdSense
and AdWords. AdSense is used by all the publishers that aim at being
part of the Google Display Network (GDN), i.e., the network of
websites that give their availability to host ads
(\textit{ad display} websites).  
%
AdWords is the service used by the advertisers to create the ads to be displayed on
the ad display websites. 
Being part of GDN
with AdSense has a monetary return for the publishers. Indeed,  after a certain number of
\textit{impressions}, the publisher obtains an amount of money as a
reward. Correspondingly, buying ad spaces with Google AdWords allows
the advertiser to generate and manage several type of ads,
including those with high personalization.
The Google patent on ``Targeted
Advertising based on user profiles and page profile"~\cite{haveliwala2012targeted} describes (at a 
high level) the way Google provides personalized advertising.

\begin{figure}[t]
\centering
\includegraphics[width=0.36\textwidth]{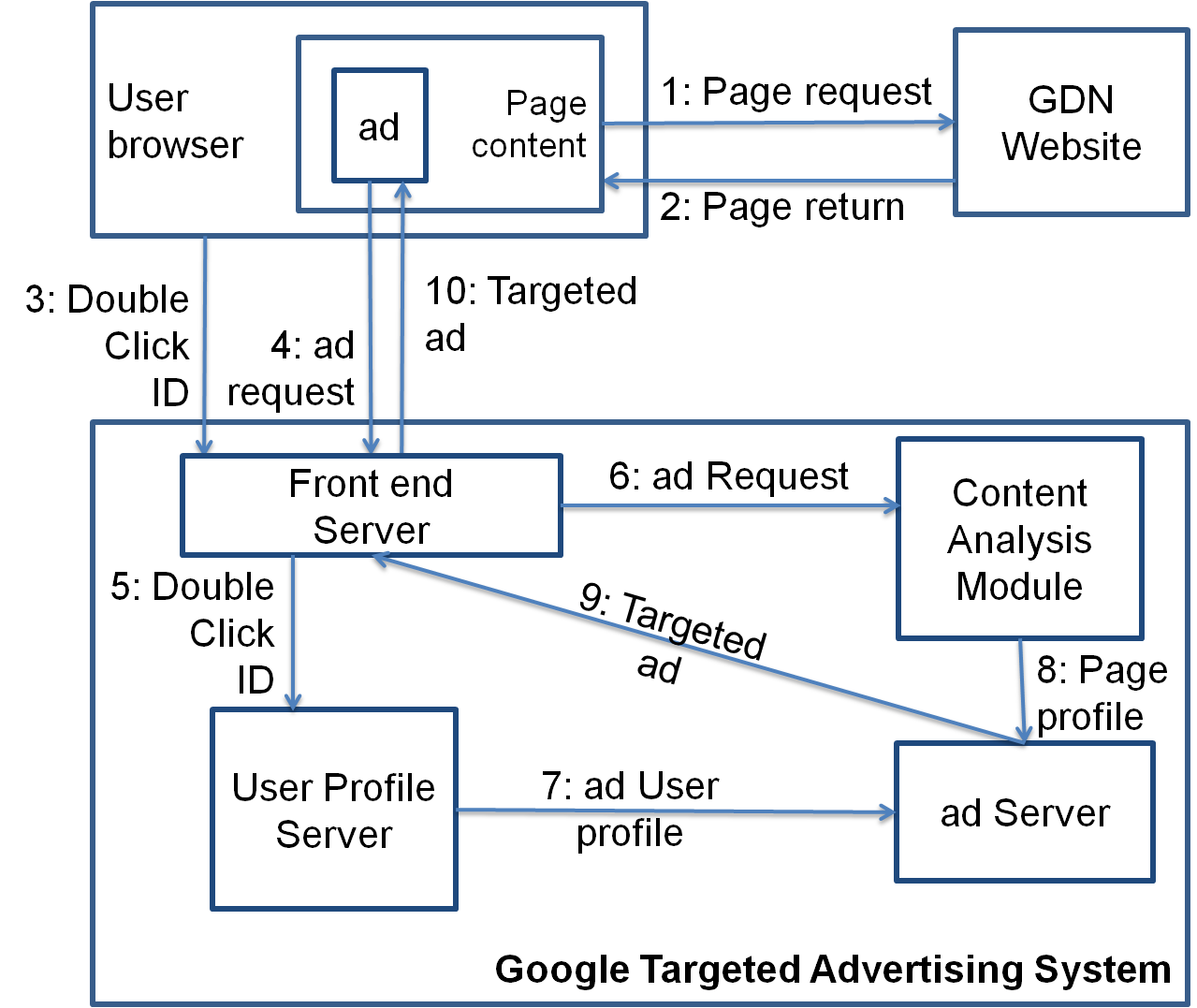}
\caption{Targeted Advertising based on user profiles and page profile\label{google}}
\end{figure} 

As illustrated in Figure~\ref{google}, when a user visits a GDN website,
she may visualize, along with the page content, both  generic and personalized advertisements.
Personalization is possible because Google monitors the GDN website
content and users' navigation behavior. In particular, for each GDN
website, a Content Analysis module evaluates the content of the GDN
web pages and builds a \textit{page profile} for each of them,
associating to each page a list of
\textit{topics}\footnote{{\it
    support.google.com/adwords/answer/156178?hl=en}}. Topics include
general categories (e.g., sport, car, art, music, and movies) as well
as sensitive and detailed information (e.g., pathologies and sexual
interests).  Then, Google builds \textit{ad user profiles}, monitoring
and learning users' behaviors when they navigate the GDN websites.
When a user visits a GDN website, her browser exchanges a DoubleClick
cookie 
tracing the navigation behavior, as described in the Google
patent~\cite{haveliwala2012targeted}. The main elements considered are
the list of the visited websites, website topics, webpage referral,
time spent on each website, number of times the user went to the
website, geo-location of the user IP address (if available),
visualized advertisements and number of clicks on them. The navigation
behavior results in the definition of the user's
\textit{interests}\footnote{{\it
    support.google.com/ads/answer/2842480?hl=en}}, which,
collectively, represent the ``ad user profile".  The patent
in~\cite{haveliwala2012targeted} describes how Google builds both page
profiles and ad user profiles.

\begin{figure}[t]
\centering
\includegraphics[width=0.36\textwidth]{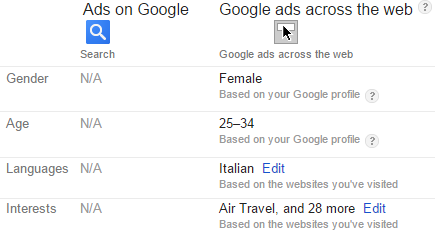}
\caption{Ad user profile for Google ads.\label{profile}}
\end{figure}

As shown in Figure~\ref{profile}, at  {\it www.google.com/settings/ads} is possible to access the ad user profile. It is worth noting that Google stores the ad user profile (gender, age, languages, and interests)  through the DoubleClick  cookie. 
Interestingly,  if the user signs in with her Google account, 
the ad profile would access gender, age and languages explicitely inserted by the user when creating her account. In Figure~\ref{profile}, the gender and the age are instead inferred by Google according to the navigation activity of an unlogged user across the Web.  

\subsection{Google AdWords: building and targeting a campaign}\label{sec:google-adwords:-buil}
AdWords makes it easy for an advertiser to create online advertisements
by means of a visual browser interface.  To configure a campaign, the
user starts specifying the ``good landing URL", i.e., the destination
URL of the user that clicks on the advertisement. Then she has to
define the advertisement to show and the keywords describing the good
(in case of search-based ads). Afterwards, it sets the
\textit{placement}, i.e., the set of websites where the ad will be
displayed (in case of display-based ads), and the time and the
geographical location the ad will appear. The next steps consist of
selecting the \textit{audience} to reach (details on the kind of
available audience are discussed below) and in setting the total
budget and the \textit{bid} (i.e., how much the advertiser will pay
for each ad impression).

The choice of the audience is a very important step, since it
determines the type of users that will see the ads.  There are several types of audiences: people who previously visited the advertiser website (\emph{in market audience}); people with specific age, gender, parental status and so on (\emph{demographic audience}); and, people interested in the specific products or services of the advertiser  (\emph{affinity audience}).
An  \emph{affinity audience} consists of aggregated users that have demonstrated a qualified interest in a certain topic. In particular, being part of an affinity audience means featuring common navigation behavior on GDN
websites with specific topics.
Examples of pre-defined affinity audiences are Cooking enthusiasts, Sports fans, Movie lovers, and many others\footnote{For a detailed list of Google affinity audiences, see: 
  {\it
   storage.googleapis.com/think/docs/affinity-audiences\_products.pdf}}.
   In addition, AdWords further refines the campaign an advertiser would like to launch, by giving her the possibility 
to reach exactly the audience she is looking for, 
specifying a list of websites users should have visited before being exposed to the campaign. Having visited such websites let users be part of \emph{custom affinity audiences}\footnote{{\it
   adwords.blogspot.it/2014/10/introducing-custom-affinity-audiences.html}}.

\begin{figure}[t]
\centering
\includegraphics[width=0.45\textwidth]{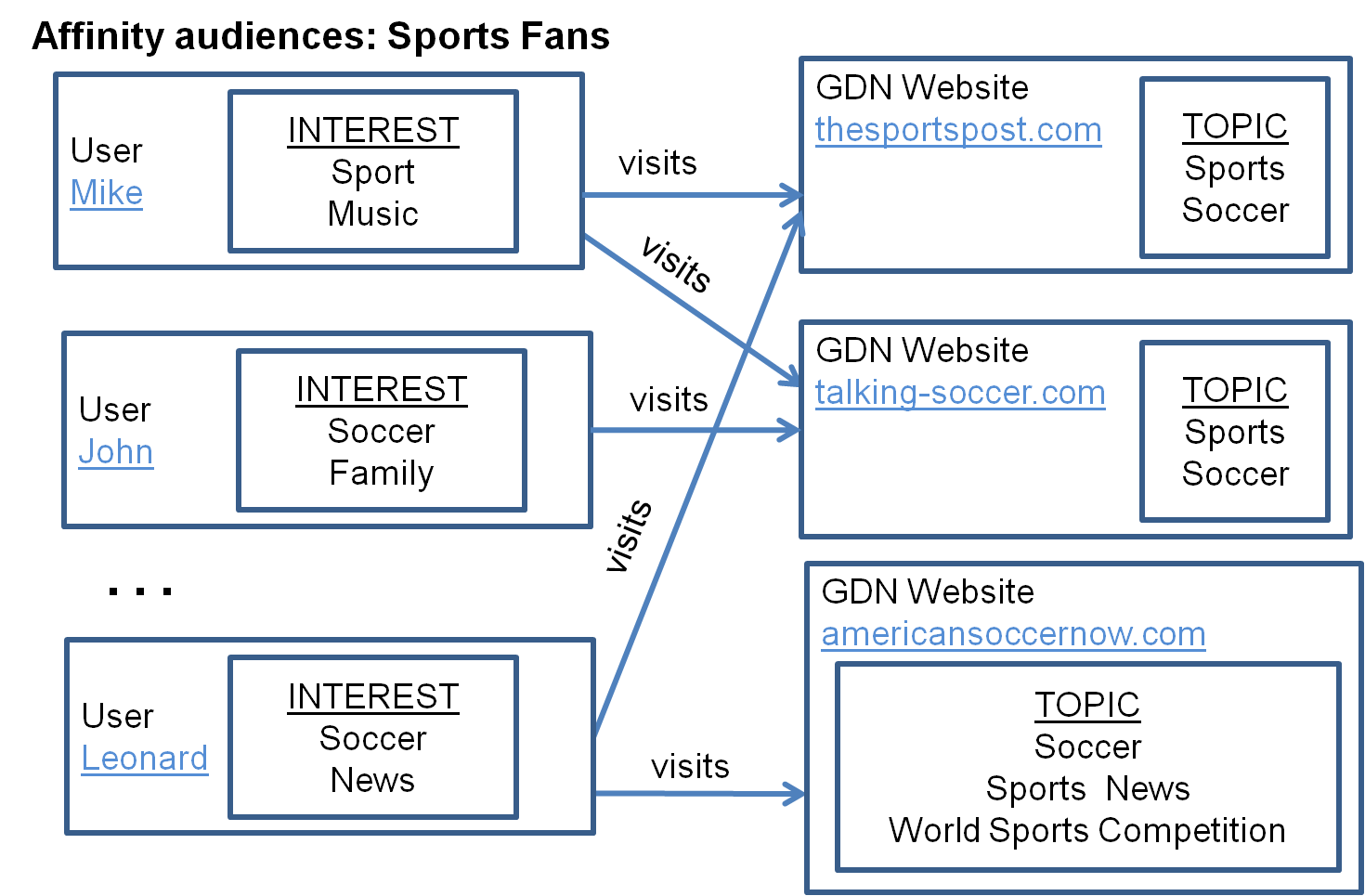}
\caption{Example of relation between topics, interests and audience\label{schema}}
\end{figure}

To represent the relation between topics,
interests and audiences, we refer
to Figure~\ref{schema}. Mike, John and Leonard belong to the affinity audience ``Sports Fans''. This is because they share a common navigation behavior through websites related to the same topics (Sports, Soccer, Sports News). Furthermore, checking the users' interests (the ones in the ads settings shown in Figure~\ref{profile}), we can see that the three users are interested in Sport and Soccer. 



Regarding a campaign budget, the advertiser needs to
choose the total budget and to set the bidding option for each ad.
AdWords provides several bidding options, depending on what matters
most to the advertiser and her business.  Advertisers can focus on
clicks, impressions, and conversions/acquisition, where acquisition
means buying a good, booking a reservation or also registering to a
mailing list.  Available AdWords bidding options are ``cost per click"
(CPC), ``cost per 1000 impressions" (CPM), ``cost per acquisition" (CPA) and
automatic bidding. Choosing  the appropriate bidding is very important,
since Google makes online auctions---based on the received bid values---%
among all the competing advertisers. Then, Google delivers the winning ad to the
user when the latter visits  the publisher page.

AdWords provides a friendly interface showing how often the ad has got an impression or a click,
and if it has reached some demographic or affinity
audience.

\section{TRAP:  targeted ads to infer ad profiles} 
\label{sec:TRAP:-our-system}

As shown in previous work (further discussed in
Section~\ref{sec:related-works}), having access to displayed ads can
lead an attacker to know users' personal information, like their
interests and their navigation behavior. Here, we go beyond the state
of the art by showing settings for and feasibility of a remote attack
capable of discovering the user affinity audience and, consequently,
inferring the information in the user ads settings. Indeed, since
affinity audiences are set according to users navigation habits, if an
attacker is able, e.g., to discover that a user is in the affinity
audience ``Job Seekers", this means that she has visited recruitment
and employment agency websites. This can lead to infer that she is
looking for a new job and raise privacy concerns (think, for example,
if such information is disclosed to the current boss of the user).




%
%

TRAP is a system for inferring user affinities of a victim user,
whenever a victim visits a website controlled by the attacker, where an advertisement controlled by the attacker is
displayed. Correlating the information from the 
website (such as the log of visiting IP addresses) and the information collected by the AdWords
system, the attacker could reconstruct information related to the (supposedly private) affinity audiences of the victim.





\subsection{Attack setup}
\label{sec:attack-setup}
\noindent To set up all the components, the attacker follows those steps:

\noindent\textbf{1) Website set up.} She sets up a website that will
be part of the GDN, registers it on AdSense and monitors the visitors'
accesses;

\noindent\textbf{2) Ad campaign set up.} She sets up an ad campaign
through AdWords, with a series of ad-hoc advertisements for several
(different) affinity audiences, which will be intentionally displayed
only on that particular GDN website;

\noindent\textbf{3) Revealing the affinity audience.} She analyzes and
compares the AdWords data with the web users' accesses, to infer the
affinity audiences and, eventually, the users' interests.

In the following, we detail each of the steps and we describe how
we have realized TRAP and the attack. 


{\bf 1) Website set up.} The first step consists in setting up a page where we can host
advertisements. For this purpose, it is enough to create and manage a
website and ask Google to add it into the GDN, in order to serve
advertisements and, eventually, to use it as controlled location of
our ads.

We have created a fake project
website (available at {\it
  www.monadsproject.com}). We have populated the website, adding consistent
and relevant contents, to make it suitable for serving ads through
AdSense, joining the GDN. A website can be part
of the GDN only when its content is 
sufficient enough for the Google crawler to automatically infer its topics.
After the AdSense analysis of the page contents, our website has been assigned to many topics,  like, e.g.,  Scientific Institutions, Scripting
Languages, and others.  We have also 
added a privacy policy explaining how we manage collected data. 
The website includes a 
landing page where the user is explicitly asked to proceed to visit
the website and share her data or not, to obtain her agreement before
acquiring and analyzing her data. Her preference is stored inside a
cookie on her browser.

Finally, to put in place the attack, there is the need to monitor in real-time the visitors activity on the website: this is fundamental to
identify the access of the victims to the website, in order to
subsequently correlate it to the shown advertisement. For this
step, several solutions are possible, like directly reading the server
log or relying on professional statistic providers. For our
experiments, we relied on
StatCounter\footnote{{\it
    www.statcounter.com}}, which, beside the log of each
website access, also provides synthetic and aggregated statistics of the
visited pages.

{\bf 2) Ad campaign set up.} The second building block consists of
setting up a personalized advertisement campaign---intentionally
limited to appear on our website. We have registered into AdWords and
we have followed standard procedures. To realize the attack, it is
important to create a campaign that relies on affinity audiences
comprising the interests we would like to extract from the victim's ad
profile. Moreover, it is critical to specify that the only website
displaying the advertisement must be ours: actually, this is the main
trick to track the audience, and, consequently, the interests of
our victims. As it will be clear in the following, if the ad is
displayed in more than one website, this perturbs the statistics on
AdWords, making the association victim--interest no more possible.

We have created a Google AdWords campaign named 
\textit{european$\_$projects}, with an advertisement group
\textit{mib$\_$project} that contains our TRAP ad, i.e., our hook for the attack. Each of the possible features of the
campaign should be carefully  set,  keeping in mind the target user, victim of the
attack. Beside the choice of the placement website, other elements allowing to restrict the range of users that will
receive an impression with our TRAP ad are, e.g.,  the geographical area,
the gender, and the age. 

The most important element to configure for the ad campaign is the affinity audience to target. AdWords will take note of
the number of impressions of our TRAP ad, for each of the audiences
set in the campaign. Being able to link each impression to a specific
user means to reveal the affinity audience of that user. 
In our
implementation, we have selected 
%
  ten affinity
audiences (among others: cooking enthusiasts, pet lovers, sports fans, music
lover, and health \& fitness buffs).

\begin{figure}[t]
\centering
\includegraphics[width=0.49\textwidth]{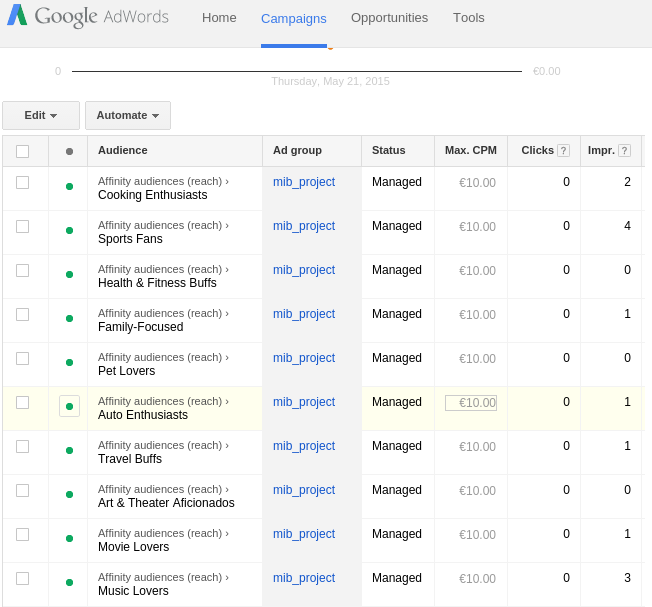}
\caption{AdWords Impressions\label{impressions}}
\end{figure}

Another important element to be set is the budget, together with the
bidding. 
Among all the advertisements that
match their affinity audiences with the user interests, Google chooses the one with
the higher bid (for example CPM, cost per 1000 impressions). Consequently, for the attack to be effective, an adequate bid
has to be set, in order to guarantee that it will be chosen among all
the competing advertisements. If the attacker wants to be really
effective, she can set a very high CPM so that, with high chances, her victims will receive the impression of her TRAP ad.  We have created an ad with a bidding based on the number of impressions, however
 the AdWords interface reports both impressions and clicks (Figure~\ref{impressions}).

%

\begin{figure}[t]
\centering
\includegraphics[width=0.45\textwidth]{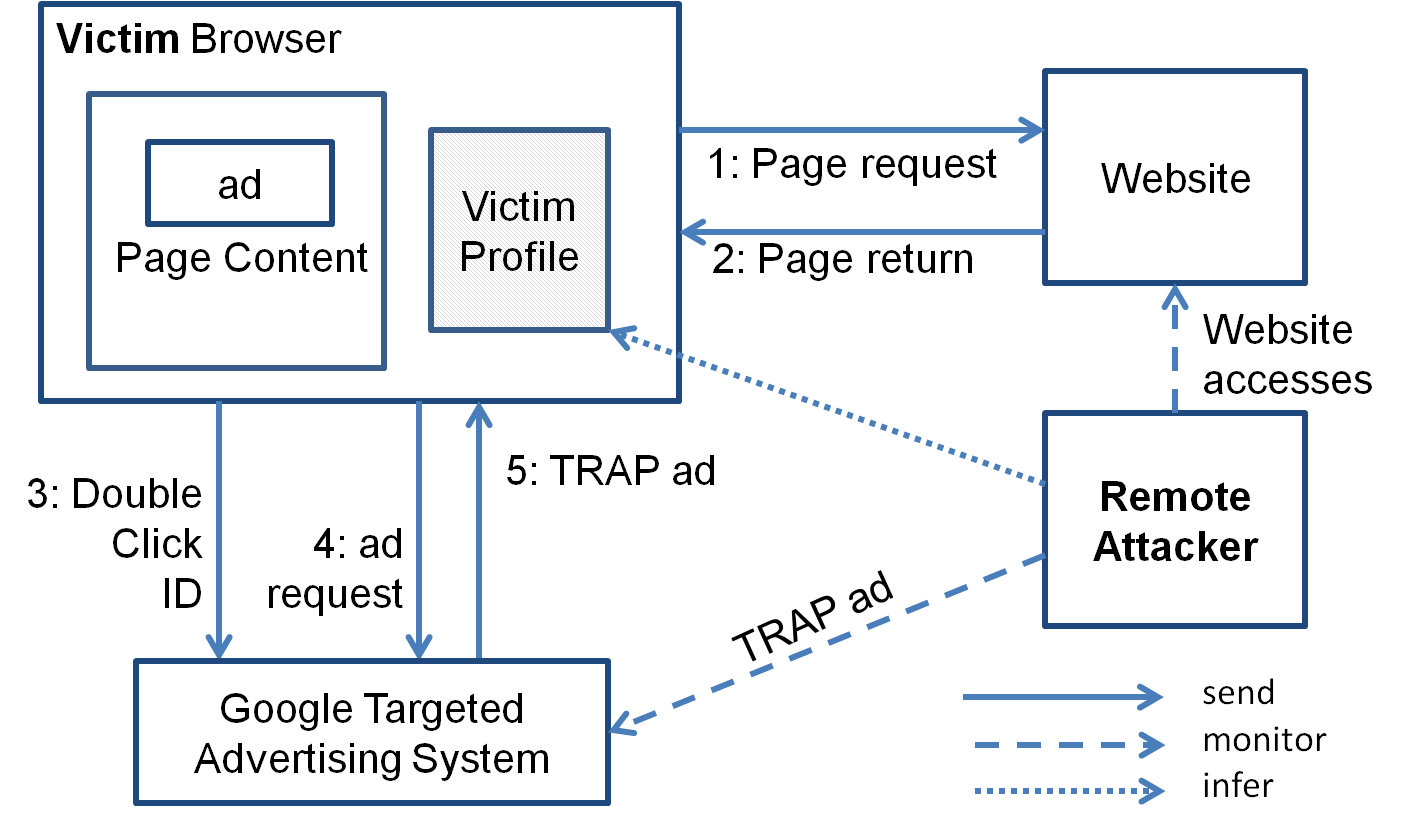}
\caption{Attack scenario\label{attack}}
\end{figure}

 {\bf 3) Revealing the affinity audience.} The third step leads to
 reveal the affinity audience of the victim.  Each time a new user
 comes onto our website and our TRAP ad receives a new impression, the
 AdWords increments the counter of the relative affinity audience
 (Figure~\ref{attack}). Moreover, since the webpages that impress the
 TRAP ad are hosted on our website, we can access the log or the
 single visits (as above described) to our page. Then, putting in
 relation the affinity audience that received a new impression with
 the visit of the user, we can associate each user to the relative
 affinity audience. This, in practice, means inferring her navigation
 behavior and, eventually, her interests.  Moreover, with these
 settings, the TRAP attack can also keep track of the users that
 clicked on our TRAP ads, not just the impressions. Analyzing the
 clicks is not considered in this work, but can be used for further
 analysis.

One limitation of this approach is that the number of impressions
and clicks received by ads are updated every 30 minutes. Then, in order to be effective, the attack has to be
suitably timed to be sure to make distinguishable the access of the
victims.  However, this limitation can be easily bypassed if the
attacker creates and monitors a single website for each of her victims: even if
it can be tedious and time consuming, it is simple and can also be
easily automatized, with a random website generator with a
database of pages with real contents.

We highlight that the TRAP attack is completely silent  and transparent for the victims. It
allows to discover users interests by only analyzing the ads Google
showed them.

\section{Attack validation}\label{sec:victim-profiles}
\begin{table*}[tb]
\begin{center}
\scriptsize
\caption{Filling ad user profiles\label{t1}}
\begin{tabular}{|p{0.4cm}|p{3cm}|p{7cm}|p{5.9cm}|}
\hline
\textbf{User}&\textbf{Visited Websites}&\textbf{Visited Websites Topics}& \textbf{User's Interests}\\
\hline
u1&\texttt{theatrehistory.com}&Arts \& Entertainment; Acting \& Theater; Literary Classics&Acting \& Theater; Broadway \& Musical Theater; History\\
\hline
u2&\texttt{carbuzz.com}&  Autos \& Vehicles; Custom \& Performance Vehicles; Vehicle Brands & Autos \& Vehicles; Performance Vehicles; Vehicle Brands\\
\hline
u3&\texttt{delish.com}&Cooking \& Recipes; Fruits \& Vegetables; Food&Beverages; Cooking \& Recipes; Fruits \& Vegetables\\
\hline
u4&\texttt{movieinsider.com}&Movies; Online Video; Movie Reviews \& Previews&Movie Reviews \& Previews\\
\hline
u5&\texttt{www.soundjay.com}&Arts \& Entertainment; Samples \& Sound Libraries; Audio Files Formats \& Codecs&Audio Files Formats \& Codecs; Samples \& Sound Libraries\\
\hline
u6&\texttt{kingpet.com}&Arts \& Entertainment; Pets; Dogs&Arts \& Entertainment; Contests, Awards \& Prizes; Dogs\\
\hline
u7&\texttt{thesportspost.com}&Sports; Team Sports; Sports News&American Football; Baseball; Fantasy Sports; Sport News\\
\hline
u8&\texttt{onetravel.com} &Travel; Air Travel; Travel Agencies \& Services& Air Travel; Travel Agencies \& Services\\
\hline
u9&\texttt{talkaboutmarriage.com}&Arts \& Entertainment, Family \& Relationships, Troubled Relationships & Family \& Relationships, marriage; Parenting, Childcare\\
\hline
u10&\texttt{youbeauty.com} \texttt{beautytask.com} \texttt{beautyandtips.com} & Beauty \& Fitness&Apparel; Bodybuilding; Cosmetology \& Beauty Professionals; Hair Care; Make-Up \& Cosmetic; Skin \& Nail Care\\
\hline
\end{tabular}
\end{center}
\end{table*}

We show the feasibility and the effectiveness of the attack by  implementing it in a controlled environment (i.e., not involving real users). In particular, we emulate ten users with ten different ad profiles. 
We let them visit one or more websites with certain topics (e.g., sport), from a web browser with clean navigation data. 
As a result, Google adds to the ad profile of each users the related interests (e.g., sport or some subcategory, such as baseball, tennis, etc.).
Table~\ref{t1} reports the websites visited by the 10 users, with the related websites topics.

We leverage
the Google display planner for selecting the websites to visit
in order to fill the profiles as we like. The Google display
planner of AdWords allows to select those GDN websites that better match a given topic.  The planner shows an ordered list of the
websites more relevant to certain interests, specified by selecting a
interest category or by simply entering keywords.  For each website, it
displays a set of charts related to the available visitor population
(age, sex, devices), together with the topic of the website and the
format of ads they display.  As an example, when selecting Sport, the
most suitable website is the first online sport magazine
{\it thesportpost.com} 
. Opening the website
details, 
it is possible to see that it
hosts ads in any format and covers the following topics: Sports, Team
Sports, Sports News.  When a visitor with a clean profile visits this
website, Google infers the following interests because of this online activity:
American Football, Baseball, Fantasy Sports, Sport News.  

\subsection{Experiment and Results}\label{sec:experiment-results}
We let the ten users (with ad profiles filled according to a specific past navigation activity) visit the Monads website, which displays the TRAP ads.  
%
%
%
For each user, we combine information gathered 
from our monitoring activity with
the data taken from AdWords. By doing so,  we are able to match a visit event with an
impression event and we infer the affinity audience of each user.  
We ultimately discover the interest of each user since we exactly know which is the TRAP ad each user sees on Monads, because a particular audience (e.g.,  \emph{Sports Fans}) is incremented by one.

In Table~\ref{t2}, we show the affinity audience that has been 
inferred for all the users under investigation.  
\setlength\tabcolsep{1.5pt}
\begin{table}
\begin{center}
\scriptsize
\caption{Real users' interests {\it vs} affinity audience inferred by impression\label{t2}}
\begin{tabular}{|l|p{5.8cm}|p{2.2cm}|}
\hline
\textbf{User}& \textbf{User's Interests}&\textbf{Affinity Audience of the \mbox{Displayed Ad}}\\
\hline
u1&Acting \& Theater; Broadway \& Musical Theater; History&Art \& Theater Aficionados\\
\hline
u2 & Autos \& Vehicles; Performance Vehicles; Vehicle Brands&Auto Enthusiasts\\
\hline
u3&Beverages; Cooking \& Recipes; Fruits \& Vegetables&Cooking Enthusiasts\\
\hline
u4&Movie Reviews \& Previews&Movie Lover\\
\hline
u5&Audio Files Formats \& Codecs; Samples \& Sound Libraries&Music Lover\\
\hline
u6&Arts \& Entertainment; Contests, Awards \& Prizes; Dogs&Pet Lover\\
\hline
u7&American Football; Baseball; Fantasy Sports; Sport News&Sports Fans\\
\hline
u8&Air Travel; Travel Agencies \& Services&Travel Buffs\\
\hline
u9& Family \& Relationships, marriage; Parenting, Childcare&Family focused\\
\hline
u10&Apparel; Bodybuilding; Cosmetology \& Beauty Professionals; Hair Care; Make-Up \& Cosmetic; Skin \& Nail Care&Health \& Fitness Buffs\\
\hline
\end{tabular}
\end{center}
\end{table}
The result of the experiment confirms that TRAP is able to discover
the users' interest through remote management and monitoring of
targeted ads.  As shown in Table~\ref{t2}, in facts, the audience
learned with our approach corresponds to the user interests or, at
least, to a subclass of them. Many associations are obvious and
self-explanatory.  User \textit{u10} needs some more explanation,
since Fitness is a subcategory of Beauty: Health and Fitness buffs are
those that care about their body and appearance.
%



\subsection{Attack optimization}

Aiming at raising the probability to attract the victim to the
attacker website, a possible optimization of our approach is the use
of personalized emails and spear phishing~\cite{hong2012state}. This
technique exploits trusted source email addresses (spoofed or not) to
invite victim users to visit some website or to pass a unique argument
when requesting a webpage (like
\textit{http://monadsproject.com?uniquearg=xyz}). Clearly, inserting a
tracking element in the email (like \textit{uniquearg=xyz} above), let
the attacker infer the victim's interests if she visits the website as
suggested in the email.  The tracking element could also be a unique
embedded image, used to monitor when the recipient accesses the
content of the email: this would require a download from the website,
unique for the target recipient.  A similar technique is used by
tracking services, like Streak\footnote{{\it
    www.streak.com/email-tracking-in-gmail}} for
Gmail.  The same approach can
be extended to a group of victims, within a set of known IP 
addresses---for example, the subnetwork of a public institute or a company. It
can also be extended to collect statistics on the interest that occurs
more within the group, leveraging TRAP ads targeted for two
complementary audiences, e.g., family focused {\it vs} travel buffs.
 

\section{Related Work}\label{sec:related-works}

In  literature,  several works deal with behavioral targeting in Online Ad Systems, highlighting 
their potential privacy threats~\cite{castellucciaSOTA}.
The novelty of the current work is that it
presents and maliciously exploits a feature of the Google advertising system,
which allows a remote attacker to infer user personal information,
as  interests and navigation behavior. In the following  we review the main research works related to privacy issues and online advertisement.

\subsubsection{Privacy violations}
In~\cite{castelluccia2012}, Castelluccia {\it et al.} show how to
reconstruct user profiles from targeted ads displayed on the users'
browser.  They provide a technique to automatically discriminate among
different kind of ads shown to the user: generic, page based, location
based, profile based. Finally, from profile-based ads they infer
users' interests.  This work is complementary to ours, since it
considers a different adversary model where the attacker is physically
behind the computer victim and sees the displayed ads in her browser.
Instead, in our scenario, the TRAP ad allows to infer the victim's
data without any access to the victim browser.

In~\cite{icdmw2010}, Korolova considers privacy violations in Facebook
through microtargeted ads.  
%
%
From publicly available information of a Facebook
account and with ad-hoc ads, the author shows that it is possible
to infer either private  information or information that the user  had configured as
\emph{Only me}, \emph{Friends only}, and \emph{Hide from these
  people}.  More broadly, one can run campaigns in order to infer 
age or gender distribution of employees of particular companies,
estimate the amount of time employees  spend on
Facebook, the fraction of employees who are interested in job opportunities
elsewhere, etc. Similarly, in~\cite{Cascavilla15, BurattinCC14}, Cascavilla et al. show several techniques to retrieve supposedly hidden information from Facebook user profile, while they do not rely on advertisements.

Targeted ads are connected with users' navigation
behavior. 
Even if users may have online interests formed and confirmed over a long time period, also short term browsing
activity can significantly impact the user's profile and, 
consequently,  change the type of ads that such user sees.  In~\cite{Meng2014}, the authors show that publishers can
subtly alter the user's profile, in order to make them the
target of the most remunerative ads. 
In~\cite{MedieKultur8070}, Bechmann highlights how profiling is related to privacy violations through a media economics and management perspective. The author analyzes different perspectives on different social media. 
Facebook focuses on user interconnections and reputation (vertical view) while Google on topic relevance (horizontal view). The work also highlights 
the implications of profiling for different stakeholders:
advertisers, developers and government agencies.
Work in~\cite{Dwyer09} analyzes the Levi's company case. The popular brand  traces users' navigation behavior through its e-commerce website,
it collects data about users and it sends them to third party websites that provide ads,  without the user's explicit consent. 
When dealing with an e-commerce website, the privacy concern is even higher: camouflaging the tracking of consumers can damage the perceived trustworthiness of the brand.

\subsubsection{Personalization}
Background and mechanisms behind the Google search engine and its
advertising system are presented in the Google patents~\cite{lawrence2005personalization} and~\cite{haveliwala2012targeted}.
While the patents describe  the big ads picture, 
they however do not reveal details about the level of
personalization in Google search and displayed ads.   Relevant personalization in query results and ads would lead to concerns about  the ``Filter Bubble" effect, where
users are trapped into information bubbles because the search
engine algorithms decide such information is relevant for them~\cite{Hannak,Pariser}.
In~\cite{Guha2010}, the authors present a measurement
methodology for determining the effectiveness of personalized online
advertising. It includes a set of guidelines for researchers that wish
to study advertising systems and an analysis of the key factors that
determine ad targeting on Google and Facebook. 


Finally, work in~\cite{Nanda2014} presents a system for personalized
web search based interest tree, a classifier able to learn users
interests and profiles from browser history. The algorithm
then reorders search results, achieving high user satisfaction. The
description of the system includes many details about the features
considered by search engines in order to determine the users'
interests and to combine them in order to understand how each feature
is relevant with respect to such interests.

\subsubsection{Privacy-preserving advertising systems}
As analyzed in~\cite{mayer12}, there exist technologies enabling the delivery of third-party 
services with affordable privacy risk. 
Some initial efforts have been
put in designing targeted advertising models yet preventing users from being
tracked by ad networks, as \emph{Privad}~\cite{Guha2009} and
\emph{Adnostic}~\cite{toubiana2010adnostic}. Their
main idea is to keep behavioral information at the client side and  to select locally the ads. Furthermore, advertisers cannot target sensitive information, like, e.g.,  health, finances, ethnicity, race, sexual orientation, personal relationships and political activity.

\section{Conclusions}\label{sec:conclusions}
\balance
With behavioral targeted advertising, the user that visits a GDN website can see
ads relevant to her real interests, inferred from her
browsing patterns. In this paper, we have seen how the technical
details that allow to realize this mechanism can be exploited by an
attacker to infer users specific interests, without directly interacting
with the users themselves. 
%
%
%
The methodology  can be exploited to target specific users by attackers
really determined in discovering their interests, clearly violating the
privacy of unaware users.  
We have also described how to extend our TRAP
system in order to overcome the slow update limitation of the AdWords
system.

While in this paper we did a preliminary assessment of the feasibility of the attack, we plan to further validate it, considering real user profiles (with the user approval). 

\section{Acknowledgments}
Mauro Conti is supported by a Marie Curie Fellowship funded by the
European Commission under the agreement PCIG11-GA-2012-321980. This
work is also supported by the Registro.it project  MIB (\emph{My Information Bubble}), the EU-India REACH project
ICI+/2014/342-896, the TENACE PRIN project 20103P34XC funded by the
Italian MIUR, and by the project ``Tackling Mobile Malware with
Innovative Machine Learning Techniques'' funded by the University of
Padua.

%
%
%
%
%
\bibliographystyle{IEEEtran}
\bibliography{arxiv-ccps}

\begin{thebibliography}{10}
\providecommand{\url}[1]{#1}
\csname url@samestyle\endcsname
\providecommand{\newblock}{\relax}
\providecommand{\bibinfo}[2]{#2}
\providecommand{\BIBentrySTDinterwordspacing}{\spaceskip=0pt\relax}
\providecommand{\BIBentryALTinterwordstretchfactor}{4}
\providecommand{\BIBentryALTinterwordspacing}{\spaceskip=\fontdimen2\font plus
\BIBentryALTinterwordstretchfactor\fontdimen3\font minus
  \fontdimen4\font\relax}
\providecommand{\BIBforeignlanguage}[2]{{%
\expandafter\ifx\csname l@#1\endcsname\relax
\typeout{** WARNING: IEEEtran.bst: No hyphenation pattern has been}%
\typeout{** loaded for the language `#1'. Using the pattern for}%
\typeout{** the default language instead.}%
\else
\language=\csname l@#1\endcsname
\fi
#2}}
\providecommand{\BIBdecl}{\relax}
\BIBdecl

\bibitem{castelluccia2012}
C.~Castelluccia, M.-A. Kaafar, and M.-D. Tran, ``{Betrayed by Your Ads!:
  Reconstructing User Profiles from Targeted Ads},'' in \emph{Proceedings of
  Privacy Enhancing Technologies}.\hskip 1em plus 0.5em minus 0.4em\relax
  Springer-Verlag, 2012, pp. 1--17.

\bibitem{icdmw2010}
A.~Korolova, ``{Privacy Violations Using Microtargeted Ads: a Case Study},'' in
  \emph{Data Mining Workshops (ICDMW)}, 2010, pp. 474--482.

\bibitem{haveliwala2012targeted}
T.~Haveliwala, G.~Jeh, and S.~Kamvar, ``Targeted advertisements based on user
  profiles and page profile,'' Nov.~27 2012, {US} Patent 8,321,278.

\bibitem{hong2012state}
J.~Hong, ``The state of phishing attacks,'' \emph{Communications of the ACM},
  vol.~55, no.~1, pp. 74--81, 2012.

\bibitem{castellucciaSOTA}
C.~Castelluccia, ``\BIBforeignlanguage{English}{{Behavioural Tracking on the
  Internet: a Technical Perspective}},'' in
  \emph{\BIBforeignlanguage{English}{European Data Protection: In Good
  Health?}}, 2012, pp. 21--33.

\bibitem{Cascavilla15}
G.~Cascavilla, M.~Conti, and I.~Y. Davide G.~Schwartz, ``{Revealing Censored
  Information Through Comments and Commenters in Online Social Networks},'' in
  \emph{IEEE/ACM Advances in Social Networks Analysis and Mining (ASONAM
  2015)}, 2015.

\bibitem{BurattinCC14}
A.~Burattin, G.~Cascavilla, and M.~Conti, ``Socialspy: Browsing (supposedly)
  hidden information in online social networks,'' \emph{CoRR}, vol.
  abs/1406.3216, 2014.

\bibitem{Meng2014}
W.~Meng, X.~Xing, A.~Sheth, U.~Weinsberg, and W.~Lee, ``{Your Online Interests:
  Pwned! A Pollution Attack Against Targeted Advertising},'' in \emph{Computer
  and Communications Security}.\hskip 1em plus 0.5em minus 0.4em\relax ACM,
  2014, pp. 129--140.

\bibitem{MedieKultur8070}
A.~Bechmann, ``{Internet profiling: the economy of data intraoperability on
  Facebook and Google},'' \emph{Journal of media and communication research},
  vol.~29, no.~55, p.~19, 2013.

\bibitem{Dwyer09}
C.~Dwyer, ``{Behavioral Targeting: A Case Study of Consumer Tracking on
  Levis.com.}'' in \emph{AMCIS}, 2009, p. 460.

\bibitem{lawrence2005personalization}
S.~Lawrence, ``{Personalization of Web Search},'' 2005, {WO} Patent App.
  PCT/US2004/030,258.

\bibitem{Hannak}
A.~Hannak, P.~Sapie, D.~Lazer, and A.~Mislove, ``{Measuring Personalization of
  Web Search},'' \emph{World Wide Web}, pp. 527--538, 2013.

\bibitem{Pariser}
E.~Pariser, \emph{{The Filter Bubble: What the Internet Is Hiding from
  You}}.\hskip 1em plus 0.5em minus 0.4em\relax {The Penguin Group}, 2011.

\bibitem{Guha2010}
S.~Guha, B.~Cheng, and P.~Francis, ``{Challenges in Measuring Online
  Advertising Systems},'' in \emph{Internet Measurement}, ser. IMC '10.\hskip
  1em plus 0.5em minus 0.4em\relax ACM, 2010, pp. 81--87.

\bibitem{Nanda2014}
A.~Nanda, R.~Omanwar, and B.~Deshpande, ``{Implicitly Learning a User Interest
  Profile for Personalization of Web Search Using Collaborative Filtering},''
  in \emph{Web Intelligence (WI) and Intelligent Agent Technologies (IAT) -
  Volume 02}.\hskip 1em plus 0.5em minus 0.4em\relax IEEE, 2014, pp. 54--62.

\bibitem{mayer12}
J.~Mayer and J.~Mitchell, ``{Third-Party Web Tracking: Policy and
  Technology},'' in \emph{Security and Privacy (SP)}, 2012, pp. 413--427.

\bibitem{Guha2009}
S.~Guha, A.~Reznichenko, K.~Tang, H.~Haddadi, and P.~Francis, ``{Serving Ads
  from localhost for Performance, Privacy, and Profit},'' in \emph{HotNets},
  October 2009.

\bibitem{toubiana2010adnostic}
V.~Toubiana, A.~Narayanan, D.~Boneh, H.~Nissenbaum, and S.~Barocas, ``Adnostic:
  Privacy preserving targeted advertising,'' in \emph{Proceedings Network and
  Distributed System Symposium}, 2010.

\end{thebibliography}
%
\end{document}